\documentclass[twocolumn]{aaph}

\usepackage{graphicx}

\usepackage{txfonts}

\usepackage{color}

\usepackage{natbib}

\begin{document}

\title{Radial and vertical epicyclic frequencies of Keplerian motion in the field of Kerr naked singularities}
 
\subtitle
{
Comparison with the black hole case\\
and possible instability of naked--singularity accretion discs
}

\author
{Gabriel T\"or\"ok\inst{\nabla}
\and
Zden{\v e}k Stuchl\'{\i}k\inst{\Delta}}

%\offprints{G.~T\"or\"ok,~ terek@volny.cz }

\institute{Institute of Physics, Faculty of Philosophy and Science, Silesian University in Opava,
Bezru{\v c}ovo n\'am. 13, CZ-746 01 Opava,
Czech Republic
\\\\e-mail:~$^\Delta$ stu10uf@fpf.slu.cz, $^\nabla$ terek@volny.cz
}

%\date{Received ... ; accepted ...}

\abstract{Relativistic Keplerian orbital frequency ($\nu_{\mathrm{K}}$) and related epicyclic frequencies (radial $\nu_{\mathrm{r}}$, vertical $\nu_{\theta}$) play an important role in physics of accretion discs orbiting Kerr black holes and can by resonant or trapping effects explain quasiperiodic oscillations observed in microquasars. Because of growing theoretical evidence on possible existence of naked singularities, we discuss behaviour of the fundamenal orbital frequencies for Keplerian motion in the field of Kerr naked singularities, primarily in order to find phenomena that could observationally distinguish a hypothetical naked singularity from black holes. Some astrophysically important consequences are sketched, namely the existence of \emph {strong resonant frequency} for all Kerr naked singularities, with radial and vertical epicyclic frequencies being equal and given by the relation
$\omega_\mathrm{sr}\,$=$\,\,a^{-2}\sqrt {a^2-1}\,(a^2+1 )^{-1}$.
   

}

\authorrunning {G. T\"or\"ok \and Z. Stuchl\'{\i}k}

\titlerunning{Radial and vertical epicyclic frequencies}

\maketitle

\section{Introduction}

Quasiperiodic oscillations (\emph{QPOs}) of X-ray brightness had been observed at {low}-(Hz) and {high}-(kHz) frequencies in some low-mass X-ray binaries  containing neutron~stars or black holes - for a review articles see, e.g., \citet{McClintockRemillard2003} in the case of black hole binaries and \citet{vanderKlis2000} in the case of  neutron~star binaries. Since the peaks of high frequencies are close to the orbital frequency of the marginally stable circular orbit representing the inner edge of Keplerian discs orbiting black holes (or neutron~stars), the strong gravity effects must be relevant in explaining high frequency QPOs \citep{AKST}. Explanation of  QPOs in the context of \emph{discs oscillations} \citep{OkazakiKatoFukue1987,NowakWagoner1991,NowakWagoner1992} has been considered for both the warped discs (trapped) oscillations \citep{KatoFukue1980,Kato2004} and resonant oscillations \citep{AbramowiczKluzniak2001,AKST}.

Usually, in the kHz QPOs the power spectrum shows twin peaks with frequencies correlated with their X-ray intensity and the peak separation being almost constant \citep[see, e.g.,][]{Strohmayer,Ford,Zhang,Klis97}.
In the case of microquasars containing stellar mass black holes, the observed ratio of the twin peak frequencies is exactly, or almost exactly, 3:2 - therefore some resonant effects are probably involved for oscillating accretion discs in microquasars \citep {KluzniakAbramowicz2000, KluzniakAbramowicz2001}\footnote{Interestingly, the same 3:2 ratio seems to be present also in the case of neutron~stars sources, indicating the same origin of the observed quasiperiodic oscillations (\citeauthor{AbramowiczBursa}, 2003; see however, \citeauthor{BelloniMendezHoman2004}, 2004).}. It was shown that the parametric resonance of the vertical and radial oscillations at epicyclic frequencies related to the Keplerian motion could be the most probable explanation of the observed microquasars phenomena \citep{TAKS}. On the other hand, the forced resonance of the epicyclic frequencies, or some other kind of resonance with ratios given by small integral numbers, e.g., 2:1, 3:1, 5:2, etc. could also explain observed QPOs frequencies (with  the same 3:2 ratio) giving them by  combinational (``beat'') frequencies \citep{AbramowiczKluzniak2001,TAKS,Aschenbach2004b}. Really, the puzzle of this 3:2 ratio kHz frequencies is still not solved definitely and other candidates to their explanation, like warped-disc oscillations \citep[see][]{Kato2004b} or simple {$p$-mode} oscillations \citep{Rezzolla2004}, are  not exluded.

Anyway, strong observational evidence supports the astrophysical relevance of disc oscillation concept introduced during eighties \citep[for a review of this concept see, e.g.,][]{BlueBook}; the mechanisms of triggering the oscillations in epicyclic frequencies were treated succefully both for thin \citep[see, e.g.,][]{BlueBook} and thick discs \citep[e.g.,][]{Matsumoto,AbramowiczKarasKluzniakLeeRebusco2003,Rezzolla2004}. Nevertheless, sophisticated three dimensional magnetohydrodynamic simulations (\emph{3-MHD}) of accretion flows, usually does not show any twin peak kHz QPOs resembling those observed \citep[][ and others]{IgumenshchevNarayanAbramowicz2003,DeVilliersHawleyKrolik2003} - only very recently \citet{KatoY} reports some view of  the 3:2 twin peaks in 3-MHD simulations. In addition, it has been shown, also very recently, by \citet{Bursa2004b} that  the possible resonant oscillations of the torus may be directly observable in X--ray modulation when occur in the inner parts of accretion flow around black hole or neutron~star (even if  the source of radiation is steady and perfectly axisymmetric).

Apparently, the vertical and radial epicyclic frequencies  of the Keplerian motion play a crucial role for both thin, Keplerian discs and thick, toroidal discs. Their properties were extensively studied in the case of accretion discs orbiting Kerr black holes in works mentioned above and in many others, and remain to be very hot outstanding topic in astrophysics of recent days. On the other hand, it is natural to extend the concept of disc oscillations in the epicyclic frequencies around other physical objects.

According to the cosmic censorship hypothesis \citep{Penrose69} and the uniqueness black-hole theorems \citep{Carter73}, the result of the gravitational collapse of a sufficiently massive rotating body is a rotating, Kerr black hole, rather than a Kerr naked singularity. Altough the cosmic censorship is a plausible hypothesis, there is some evidence against it.
Naked singularities arise in various models of spherically symmetric collapse \citep[e.g,][]{Lake1990,Joshi2002,Joshi2004}. In modelling the collapse of rotating stars, it had been shown that in some situations the mass shedding and gravitational radiation will not reduce the angular momentum of the star enough to lead to the formation of a Kerr black hole \citep{MillerFelice}. Candidates for the formation of Kerr naked singularities with a ring singularity were found in scenario of  \citet{CharltonClarke} and some 2D numerical models of collapsing, rotating supermassive objects imply that a Kerr--like naked singularity could develop from objects rotating rapidly enough \citep{Nakamura}. 

It is generaly believed that black holes are stable against perturbations that would transfer them into naked singularities \citep{Bardeen1973,Thorne1974,Wald1974,CohenGautreau1979,Israel1986,FeliceYu1986}. However, recently presented gedanken experiments concerning electrically charged, Reissner--Nordstr\"{o}m black holes put some doubts on this believe. It was shown that a charged test particle radially falling into a nearly Reissner--Nordstr\"{o}m black hole may transfer it into a Reissner--Nordstr\"{o}m naked singularity (\citeauthor{Hubeny1999}, 1999; see however, \citeauthor{QuinnWald1999}, 1999 for a more detailed analysis).  Further, it was shown that an extreme Reissner--Nordstr\"{o}m black hole could be transfered into a Kerr--Newman naked singularity by capturing a flat and electrically neutral spinning body that plunges in radially with its spin aligned to the radial direction \citep{FeliceYu2001}. Moreover, possible existence of naked singularities is supported by general mathematical studies concerning scalar fields around Reissner--Nordstr\"{o}m naked singularities \citep[see, e.g., ][]{Stalker2004}. 

Because the cosmic censorship hypothesis is far from being proved, naked--singularity spacetimes related to the black--hole spacetimes with a nonzero charge and/or rotation parameter could be considered conceivable models for some exotic Galactic binary systems or, on much higher scale, of quasars and active galactic nuclei, and deserve some attention. Of particular interest are those effects that could observationally distinguish a naked singularity from black holes. Therefore, we shall discuss here in detail the properties of the vertical and radial epicyclic frequencies of the Keplerian circular motion in the field of Kerr naked singularities in order to find astrophysically relevant differencies between the black-hole and naked-singularity cases. Indeed, we will show that in the Kerr naked singularity spacetimes a new, and astrophysically very important effect exist, which does not occur in the case of black holes. Namely, for any Kerr naked singularity a stable circular geodesics exist on which the vertical and radial epicyclic frequencies are equal indicating thus possibility of an extremely strong resonance and instability of the Keplerian accretion discs.

In Sect. \ref{Formulae} we present the well known formulae giving the vertical and radial epicyclic frequencies of the Keplerian circular orbits in the Kerr spacetimes. In Sect. \ref{Properties} we discuss properties of these frequencies for both black-hole and naked-singularity spacetimes. We focus attention on the extrema of both the radial  and vertical frequencies, if they exist. In sect. \ref{Implications}, the naked-singularity case is discussed in detail and the strong resonant frequency, when the epicyclic frequencies are equal, is given as a function of the naked singularity rotational parameter. In Sect. \ref{conclusions} we present some concluding remarks.
  
\section{Epicyclic oscillations of Keplerian discs}
\label{Formulae}

%%%%%%%%%%%%%%%%%%%%%%%%%%%%%%%%
%%%%   Figure KeplerianVsEpicyclic  %%%%%%%%%%%%%
%%%%%%%%%%%%%%%%%%%%%%%%%%%%%%%%
\begin{figure*}
\resizebox{1\hsize}{!}{\includegraphics{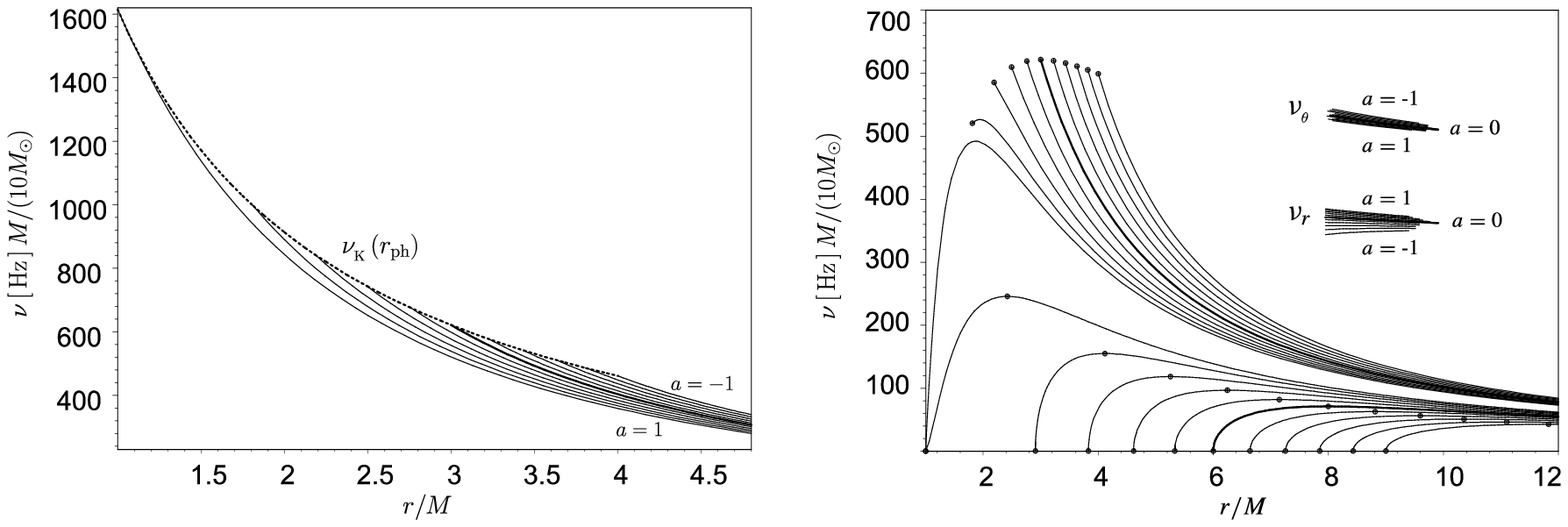}}
\caption{ Keplerian frequency $\nu_K$ (left panel - \emph{taken from \citeauthor{AKST}, 2004}), and the two
epicyclic frequencies (right panel) for
Keplerian circular orbits around Kerr black holes. The curves are spaced by 0.2 in $a$.
%\hfill\emph{We note that except the right lower figure this block of plots is taken from AKST.}
} 
\label{KeplerianVsEpicyclic}
\end{figure*}
%%%%%%%%%%%%%%%%%%%%%%%%%%%%%%%
%%%%%%%%%%%%%%%%%%%%%%%%%%%%%%%

\noindent
In the case of oscillating Keplerian disc three orbital frequencies are relevant:
Keplerian orbital frequency $\nu_{\mathrm{K}} =\Omega_{\mathrm{K}}/2\pi$,
radial epicyclic frequency $\nu_{r}\,$=$\,\omega_{r}/2\pi$,
and vertical epicyclic frequency $\nu_{\theta}\,$=$\,\omega_{\theta}/2\pi$. 
For discs orbiting Kerr black holes or naked singularities corresponding angular velocities $\Omega_{\mathrm{K}}$, $\omega_\mathrm{r}$, $\omega_{\theta}$ are given
\footnote{We use Boyer-Lindquist coordinates,
$\left( t, r,\theta, \phi \right)$. We rescale the central object mass with
$M = GM_0/c^2 = r_{\rm G}$ and the central object angular momentum with
$a = J_0c/\mathrm{G}M_0^2$. Here, the parameters $M_0$ and $J_0$ give the mass and the internal angular momentum of the Kerr black hole or naked singularity.}
by the well known formulae \citep[e.g.,][]{NowakLehr1999},
\smallskip
\begin{eqnarray}
\label{Keplerian}
\Omega_{\mathrm{K}}&=&\left ({{GM_0}\over {r_G^{~3}}}\right )^{1/2}\left( x^{3/2} + a \right)^{-1},
\\
\nonumber
\\
\label{radial}
\omega_\mathrm{r}^2 &=&\alpha_\mathrm{r}\,\Omega_\mathrm{K}^2,
\\
\label{vertical}
\omega_{\theta}^2 &=&\alpha_\theta\,\Omega_\mathrm{K}^2,
\end{eqnarray}
where 
\begin{eqnarray}
\alpha_\mathrm{r}\left( x\,,a\right)&\equiv&{1-6\,x^{-1}+ 8 \,a \, x^{-3/2} -3 \, a^2 \, x^{-2}},
\\
\alpha_\theta\left( x\,,a\right)&\equiv&{1-4\,a\,x^{-3/2}+3a^2\,x^{-2}},
\end{eqnarray}
and x is introduced as dimensionless radial coordinate
\begin{equation}
\label{x}
x=r/M.
\end{equation}
In the limit of the Schwarzschild black holes ($a=0$), we arrive at
\begin{eqnarray}
\alpha_\mathrm{r}\,(x)&=&1-\frac{6}{x},
\\
\alpha_\theta\,(x)&=&1\,,
\end{eqnarray}
so that $\Omega_{\mathrm{K}}\,(x)\,$=$\,\omega_\theta\,(x)$
.%(Fig. \ref{SchwarzschildVsKerr} - left panel).
\\
In the field of Kerr black holes ($a\,$$\neq\,$0), there is
\begin{equation}
\Omega_{\mathrm{K}}\,(x,\,a)>\omega_\theta\,(x,\,a)>\omega_\mathrm{r}\,(x,\,a)
\end{equation}
in the range where the frequencies are well defined (Fig. \ref{KeplerianVsEpicyclic}~-~right panel).
\medskip

\noindent
The properties of $\Omega_{\mathrm{K}},$ $\omega_\mathrm{r},$ $\omega_\theta$ for Kerr black--hole spacetimes are rewieved, e.g., in \citet{BlueBook}.
We can summarize that
\begin{itemize}
\item
the Keplerian frequency 
is monotonically decreasing function of radius for all the range of black hole rotational parameter $a\in (-1,\,1)$
\footnote{Here and henceforth values of $a\,$$>$$\,0$ correspond to corotating orbits, while $a\,$$<$$\,0$ give counterrotating orbits.}
~in astrophysically relevant radii above the photon orbit (Fig. \ref{KeplerianVsEpicyclic} - left panel)
\item
for slowly rotating black holes the vertical epicyclic frequency is monotonically decreasing function of  radius in the same radial range as well, however, for rapidly rotating black holes this function  has a maximum (Fig. \ref{KeplerianVsEpicyclic} - right panel)
\item the radial epicyclic frequency 
has a local maximum for all $a\in (-1,\,1)$ (Fig. \ref{KeplerianVsEpicyclic} - right panel)
\end{itemize}

For Kerr naked singularities the behaviour of the epicyclic frequencies is different.
We show in the next sections that the vertical frequency can has two local extrema and the radial one even three local extrema. For completeness, we shall give detailed discussion of the properties of the functions $\Omega_\mathrm{K}(x,\,a)$, $\omega_\mathrm{r}(x,\,a)$, and $\omega_\theta(x,\,a)$ for both naked singularities and black holes.
\medskip

Obviously, all three frequencies (\ref{Keplerian})-(\ref{vertical}) have the general form,
\begin{equation}
\nu=\left ({{GM_0}\over {r_G^{~3}}}\right )^{1/2}\,f\,(x,\,a)\doteq 32.3\left(\frac{M_0}{M_\odot}\right)\,f\,(x,\,a)\,\mathrm{kHz}.
\end{equation}
Thus for the reader's convenience we express the frequency as $\nu\,$[Hz]$\,M/(10M_\odot)$ in every quantitative plot of frequency dependence on radial coordinate (\ref{x}), i.e., displayed $value$ is the frequency relevant for central object with mass of 10$M_\odot$, which could be simply rescaled for another mass just dividing the displayed $value$ by the respective mass in units of ten solar mass.

\section{Properties of the epicyclic frequencies}
\label{Properties}

First, it is important to find the range of relevance of the functions $\Omega_\mathrm{K}(x,\,a)$, $\omega_\mathrm{r}(x,\,a)$, and $\omega_\theta(x,\,a)$
above the event horizon located at
\begin{equation}
x_{+}=1+\sqrt{1-a^2}
\end{equation}
for black holes, and above the ring singularity located at
\begin{equation}
x=0 \quad (\theta=\pi/2)
\end{equation}
for naked singularities.

The circular geodesics in the field of  Kerr black holes were discussed in \citet{Bardeen1972}, while in the case of Kerr naked singularities  the circular geodesics were discussed in \citet{Stuchlik1980}. 
\smallskip

\noindent
We can summarize that circular geodesics can exist in the range of
\begin{equation}
x\in\left( x_\mathrm{ph}\,(a), \infty\right),
\end{equation} 
where 
\begin{equation}
x_\mathrm{ph}\,(a)=
2\,
\left [
1+\mathrm{cos}
\left(
\frac{2}{3}\,\mathrm{arccos}(-a)
\right)\,
\right ]
\end{equation} 
gives loci of photon circular geodesics. Stable circular geodesics, relevant for the Keplerian, thin accretion discs exist in the range of
\begin{equation}
x\in\left( x_\mathrm{ms}\,(a), \infty\right),
\end{equation}
where $x_\mathrm{ms}\,(a)$ denotes radius of the marginally stable orbit, determined (in an implicit form) by the relation
\begin{equation}
{1-6\,x^{-1}+ 8 \,a \, x^{-3/2} -3 \, a^2 \, x^{-2}}=0,
\end{equation}
which coincides with the condition
\begin{equation}
\alpha_\mathrm{r}\,(x,\,a)=0.
\end{equation}
For toroidal, thick accretion discs the unstable circular geodesics can be relevant in the range
\begin{equation}
x_\mathrm{mb}\leq x_{\mathrm{in}}<x<x_\mathrm{ms},
\end{equation}
being stabilized by pressure gradients in the tori. Here,
\begin{equation}
x_\mathrm{mb}=2-a+2\sqrt{1-a}
\end{equation}
is the radius of the marginally bound circular geodesic that is the lower
limit for the inner edge of thick discs \citep{KozlowskiJaroszynskiAbramowicz1978, KrolikHawley2002}. 
\smallskip

Clearly, the Keplerian orbital frequency is well defined up to $x\,$=$\,x_\mathrm{ph}
\,(a)$. However, $\omega_\mathrm{r}$ is well defined, if $\alpha_\mathrm{r}\,$$\geq$$\,0$, i.e., at  $x\,$$\geq$$\,x_\mathrm{ms}\,(a)$, and $\omega_\mathrm{r}\,(x)\,$=$\,0$ at $\,x_\mathrm{ms}$.
We can also show that for $x\,$$\geq$$\,x_\mathrm{ph}$, there is $\alpha_\theta\,$$\geq$$\,0$, i.e., the vertical frequency $\omega_\theta$ is well defined at $x\,$$>$$\,x_\mathrm{ph}$.

%%%%%%%%%%%%%%%%%%%%%%%%%%%%%%

\subsection{Local extrema of  epicyclic frequencies}

%%%%%%%%%%%%%%%%%%%%%%%%%%%%%%%%
%%%%   Figure RadialExtrems  %%%%%%%%%%%%%
%%%%%%%%%%%%%%%%%%%%%%%%%%%%%%%%
\begin{figure*}[!]
\begin{minipage}{1\hsize}
{\includegraphics[width=\hsize]{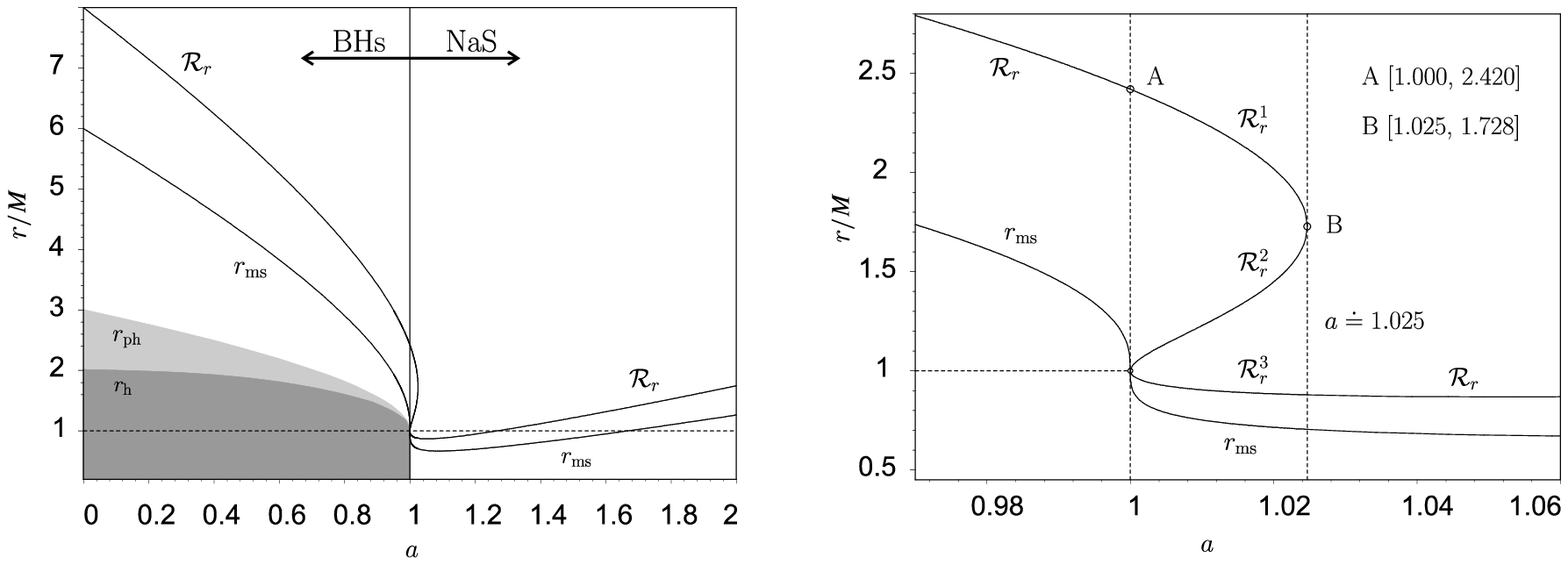}}
\end{minipage}
\caption{The locations $\mathcal{R}^{\,i}_\mathrm{r}$ of   the radial epicyclic frequency local extrema. On right panel detailed view is shown. [Here, in the next Fig. and hencefort we use the following convention for both kind of extrema of the radial ($\mathcal{R}^{\,i}_\mathrm{r}$) and the vertical ($\mathcal{R}^{\,i}_\theta$) epicyclic frequencies: odd or missing superscript denotes a local maximum and even-numbered one means a local minimum]. For the answer to question whether in the case of naked singularities the maximum is global one see left panel in Fig. \ref{DegeneracyAndRatio}.}
\label{RadialExtrems}
\end{figure*}
%%%%%%%%%%%%%%%%%%%%%%%%%%%%%%%%

Denoting  by $\mathcal{R}_{\mathrm{K}}$,  $\mathcal{R}_{r}$, $\mathcal{R}_{\theta}$ the local extrema of Keplerian $\nu_\mathrm{K}$ and epicyclic $\nu_\mathrm{r}$, $\nu_\theta$ frequencies,  we can give the extrema by the condition 
\begin{equation}
\label{extremscondition}
\frac{\partial}{\partial\,r}\,\nu_\mathrm{i}=0
\Leftrightarrow \frac{\partial}{\partial\,x}\,\nu_\mathrm{i}=0
\quad\mathrm{for~~}\mathcal{R}_\mathrm{i}, \quad\quad \mathrm{i}\in\{\mathrm{K},r,\theta\},
\end{equation}
where $x$ is dimensionless coordinate (\ref{x}).
From (\ref{Keplerian})-(\ref{vertical}), we find that the corresponding derivatives
\footnote{After introducing $^\prime$ as $\mathrm{d}/\mathrm{d}r$.}
are
\begin{eqnarray}
\label{KeplerianDerivativ}
\Omega_{\mathrm{K}}^{\,\prime}&=&-\frac{3}{2}\,\sqrt {{\frac {GM_{{0}}}{{r_{{G}}}^{3}}}}\,\frac{\sqrt {x}}{\left( {x}^{3
/2}+a \right) ^{2}},
\\
\label{EpicyclicDerivativ}
\omega_{\mathrm{j}}^{\,\prime}&=&\phantom{-}\frac{3}{2}\,\left[\frac{2\,\beta_{\mathrm{j}}}{\sqrt\alpha_{\mathrm{j}}}-\frac{\sqrt{\alpha_\mathrm{j}\,x}}{\left({x}^{3
/2}+a\right)}\right]\,\Omega_\mathrm{K},
\\
\alpha_\mathrm{j}^{\,\prime}&=&\phantom{-}6\,\beta_\mathrm{j}}/{\alpha_\mathrm{j},
\end{eqnarray}
where
~$\mathrm{j}\in\{r,\theta\},\quad$ and
\begin{equation}
\beta_{r}\,(x,\,a)=\frac{1}{x^{2}}-2\,\frac {a}{x^{5/2}}+ \frac {a^2}
{x^3},
\end{equation}
\begin{equation}
\beta_{\theta}\,(x,\,a)=\frac {a}{x^{5/2}}-\frac {a^2}{x^3}.
\end{equation}
Clearly, $\Omega_{\mathrm{K}}^{\,\prime}\,$$<$$\,0$ for $x\,$$>$$\,0$, i.e., the Keplerian frequency is a monotonically decreasing function of  the radial coordinate for any value of the rotational parameter $a$.
\smallskip

Relations (\ref{extremscondition}) and (\ref{EpicyclicDerivativ}) imply the condition determining extrema $\mathcal{R}_\mathrm{j}\,(a)$ of the epicyclic frequencies:
\begin{equation}
\label{implicitcondition}
\beta_\mathrm{j}\,(x,\,a)=\frac{1}{2}\frac{\sqrt{x}}{x^{3/2}+a}\,\alpha_\mathrm{j}\,(x,\,a)\quad\quad \mathrm{j}\in\{r,\theta\}.
\end{equation}

\begin{figure*}[!]
\begin{minipage}{1\hsize}
{\includegraphics[width=\hsize]{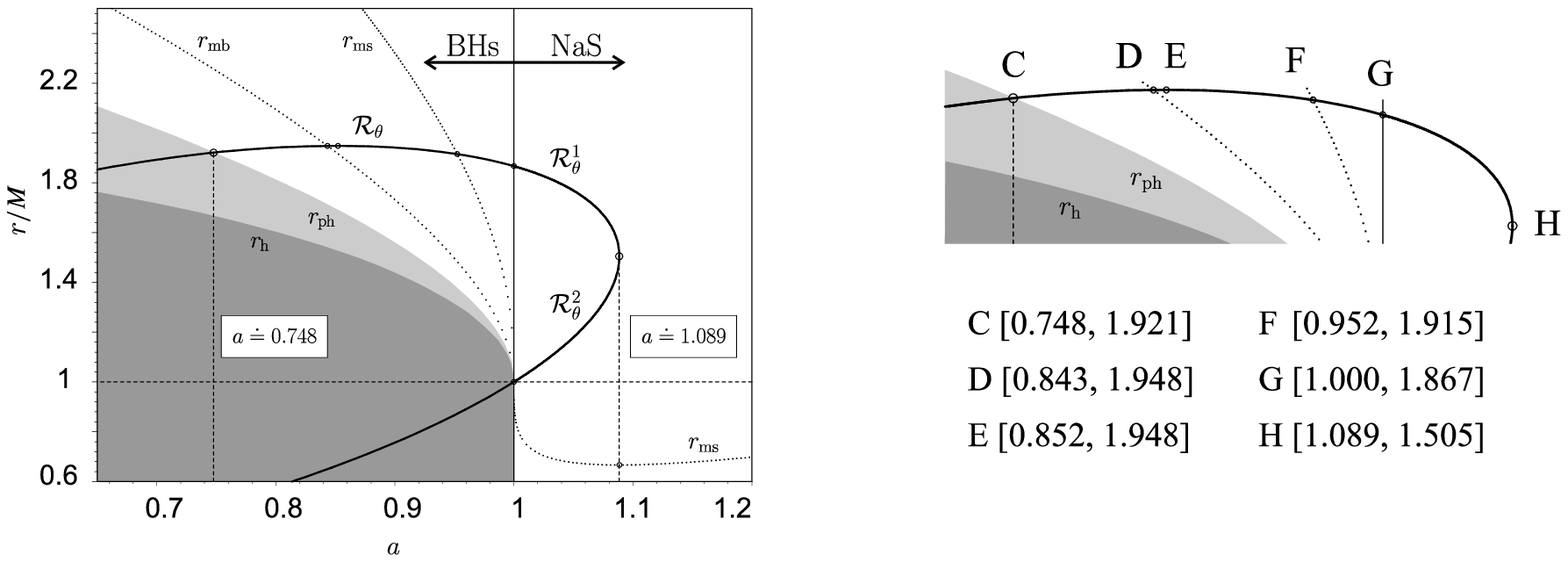}}
\end{minipage}
\caption{
The locations $\mathcal{R}^{\,i}_\theta$ of  the vertical epicyclic frequency local extrema.
The right panel gives an exact information about position of the important points.
%\vspace{.6cm}
}
\label{VerticalExtrems}
\end{figure*}

\noindent
Because we have checked that in the case of  counterrotating orbits ($a$$\,<\,$0) the extrema $\mathcal{R}_\theta$ are located under the photon circular orbit and the extrema $\mathcal{R}_\mathrm{r}$ are just extension of  the $\mathcal{R}_\mathrm{r}$ for corotating case ($a$$\,<\,$0), we are focused mainly to the case of corotating orbits in the next discussion.
In Figs \ref{RadialExtrems} (\ref{VerticalExtrems}) we show curves~~$\mathcal{R}_\mathrm{r}^{\phantom{r}\mathrm{k}}\,(a)$, $\mathrm{k}\in\{1,2,3\}$ $\left(\,\mathcal{R}_\theta^{\phantom{\theta}\mathrm{l}}\,(a),~\mathrm{l}\in\{1,2\}\right)$ implicitly determined by the relations (\ref{implicitcondition}); indexes k, l denote different branches of the solution of  (\ref{implicitcondition}).
\medskip

\begin{figure*}[!]
\begin{minipage}{1\hsize}
{\includegraphics[width=\hsize]{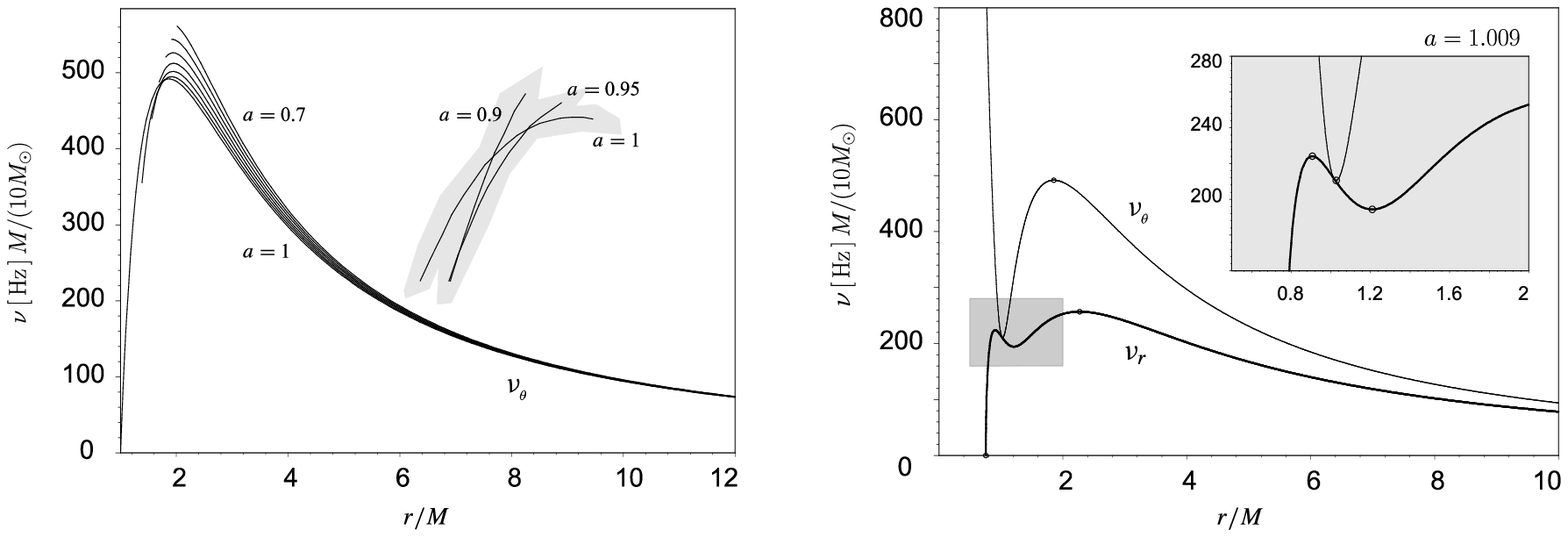}}
\end{minipage}
\caption{Left panel illustrate ``unlikely`` effect resulting from the existence maxima of  $\mathcal{R}_{\theta}$ (the point E on Fig. \ref{VerticalExtrems}). Curves $\nu_\theta\,(r)$ after $a\doteq 0.852$ cross each other (curves \emph{differ} in rotational parameter here by 0.05), see also Fig. \ref{KeplerianVsEpicyclic} for comparison.
Right panel displays an example of  epicyclic frequency behaviour for Kerr naked singularity with $a=1.009$; all allowed extrema are present. Note that the minimum $\nu_\theta$ is very close but not identical with the \emph{point of contact} (which is also present - see Sect. \ref{StrongFrequency} for details).
%\vspace{-.7cm}
}
\label{CrossingAndExample}
\end{figure*}

\noindent
The radial epicyclic frequency has one local maximum for Kerr black holes
\begin{equation}
-1\leq a \leq 1,
\end{equation}
but it  has two local maxima
\footnote{We distinguish which is the global one in Sect. \ref{Implications}.}
and one local minimum for Kerr naked singularities with
\begin{equation}
1<a<a_{\mathrm{c}(r)}\doteq 1.025,
\end{equation}
and again one local maximum for
\begin{equation}
a \geq a_{\mathrm{c}(r)} \quad \mathrm{and} \quad a<-1.
\end{equation}
The vertical epicyclic frequency has a local maximum at $x\,$$>$$\,x_\mathrm{ph}$ for Kerr black holes with
\begin{equation}
a > a_{\mathrm{ph}\,(\theta)} \doteq 0.748,
\end{equation}
and at $x\,$$>$$\,x_\mathrm{ms}$ for 
\begin{equation}
a > a_{\mathrm{ms}\,({\theta})} \doteq 0.952.
\end{equation}
The local maximum of  $\omega_{\theta}\,(x,\,a)$ is relevant in resonant effects for $a\,$$>$$\,a_{\mathrm{ms}\,(\theta)}$.  
Note that $\mathcal{R}_{\theta}$ has a maximum at
\begin{equation}
a_{\mathrm{max}\,\left({\mathcal{R}_\theta}\right)}\doteq 0.852,
\end{equation}
therefore, the situation with function $\omega_\theta\,(x,\,a)$ is more complicated than it seems to be indicated by Fig. \ref{KeplerianVsEpicyclic}: for high values of the black hole rotational parameter $a$, curves $\omega_\theta\,(x,\,a)$ cross each other as is shown in left panel of Fig. \ref{CrossingAndExample} - Fig. \ref{KeplerianVsEpicyclic} does not show such detail because of  hi-spacing between curves.\\ 
In the Kerr naked singularity spacetimes, the function $\omega_\theta\,(x,\,a)$ has a local minimum and a local maximum for
\begin{equation}
1< a < a_{\mathrm{c}{_\theta}} \doteq 1.089,
\end{equation}
and it has no astrophysically relevant local extrema for 
\begin{equation}
a \geq a_{\mathrm{c}_\theta} \quad \mathrm{and} \quad a<-1.
\end{equation}

%%%%%%%%%%%%%%%%%%%%%%%%%%%%%%%%
%%%%   Figure VerticalExtrems  %%%%%%%%%%%%%
%%%%%%%%%%%%%%%%%%%%%%%%%%%%%%%%

Using properties of  $\mathcal{R}_\mathrm{r}\,(a)$ and $\mathcal{R}_\theta\,(a)$, we can conclude that there exist two qualitatively different types of behaviour of the epicyclic frequencies in the Kerr black-hole spacetimes and three qualitatively different types of their behaviour in the Kerr naked-singularity spacetimes. Examples of the behaviour of the epicyclic frequencies for Kerr black holes are given in Fig. \ref{KeplerianVsEpicyclic} (see also Fig. \ref{SummaryBH}).

An example of the behaviour of the epicyclic frequencies in Kerr naked--singularity spacetimes is shown in Fig. \ref{CrossingAndExample} (right panel) for the case when all the local extrema mentioned above are present, while for an example of the case  when the number of the local extrema is lowest see Fig. \ref{ExampleAndLocation}. The complete set of figures representing systematically the evolution of the character of the epicyclic frequencies with rotational parameter growing is included in the Appendix which consists of  Figs \ref{SummaryBH} (black holes) and \ref{SummaryNaS} (naked singularities)  and fold also the evolution of  derivatives (\ref{EpicyclicDerivativ}) and of the ratio $\nu_\theta/\nu_\mathrm{r}$ of the epicyclic frequencies. 
This set of figures represents classification of the Kerr spacetimes according to the properties of the epicyclic frequencies which is fully given in the Sect. \ref{conclusions}. Note that in the black-hole case it is important to distinguish the cases when the local maximum of  $\nu_\theta\,(x,\,a)$ is located above $x_\mathrm{ms}$, and under $x_\mathrm{ms}$.

Clearly, the behaviour of the epicyclic frequencies substantially differs for Kerr naked singularities in comparison with  Kerr black holes. We discuss some consequences of this different behaviour for naked singularities in Sect. \ref{Implications}.

Now, we focus our attention on some properties of the epicyclic frequencies that are very important for treating the resonant oscillation phenomena from the observational point of view.

%%%%%%%%%%%%%%%%%%%%%%%%%%%%%%%%
%%%%   Figure   CrossingVsExample   %%%%%%%%%%%%%
%%%%%%%%%%%%%%%%%%%%%%%%%%%%%%%%

\subsection{Ratio of epicyclic frequencies}

The ratio of epicyclic frequencies $\nu_\theta$ and $\nu_\mathrm{r}$ is important in a well defined way for some models of QPOs \citep[e.g.,][]{AKST, Kato2004}.
\\
It is well known \citep[see, e.g.,][]{BlueBook} that for the Kerr black holes (-1$\,\leq\,$$a$$\,\leq\,$$1$) the inequality
\begin{equation}
\omega_\mathrm{r}\,(x,\,a)<\omega_\theta\,(x,\,a)
\end{equation} 
holds, i.e., the equation
\begin{equation}
\omega_\mathrm{r}\,(x,\,a)=\omega_\theta\,(x,\,a)
\end{equation}
does not have any real solution in all range of black hole rotational parameter $a\in\left(-1,\,1\right)$ and
\begin{equation}
\left(\frac{\nu_\theta} {\nu_\mathrm{r}}\right)> 1
\end{equation}
for any Kerr black hole. Furthermore, this ratio is a monotonic function of radius for any fixed $a\in\left(-1,\,1\right)$ (see Fig. \ref{Ratios} - left panel). However, the situation is different for Kerr naked singularities, see Sect. \ref{StrongFrequency}. 

\subsection{Implications for the orbital resonance models in the field of Kerr black holes}

%%%%%%%%%%%%%%%%%%%%%%%%%%%%%%%%
%%%%   Figure   Ratios   %%%%%%%%%%%%%
%%%%%%%%%%%%%%%%%%%%%%%%%%%%%%%%
\begin{figure*}[ht!]
\begin{minipage}{1\hsize}
{\includegraphics[width=\hsize]{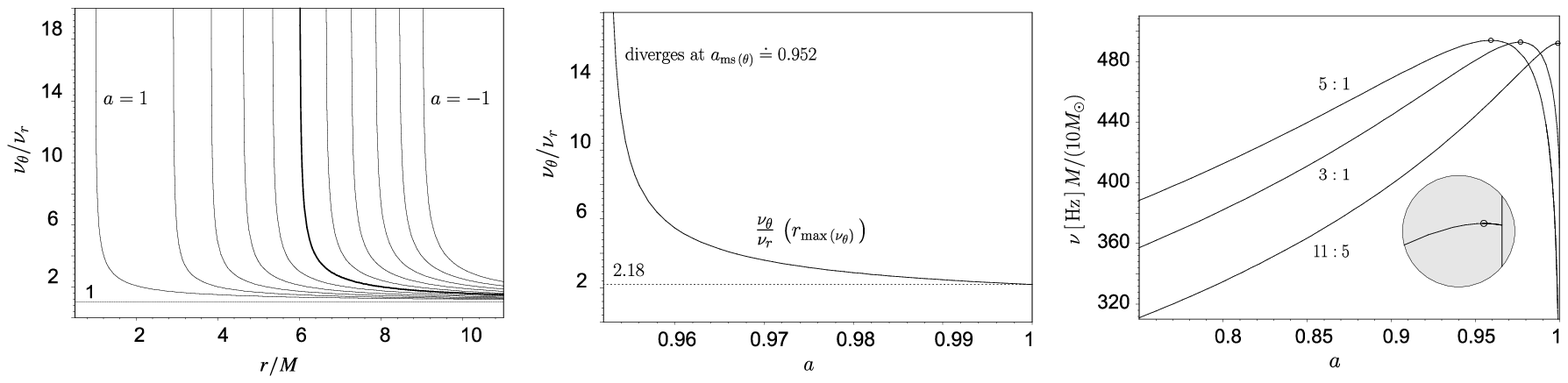}}
\end{minipage}
\caption{Left panel shows the behaviour of  ratio $\nu_\theta/\nu_\mathrm{r}$ of the epicyclic frequencies - curves are spaced by 0.2 in rotational parameter. Middle panel gives the information about ratio between epicyclic frequencies at maxima of $\nu_\theta$, while right panel illustrates the examples of behaviour of the frequency $\nu_\theta$ for three cases of forced resonances with $p/q\,$= 5, 3, and 2.2. 
}
\label{Ratios}
\end{figure*}
%%%%%%%%%%%%%%%%%%%%%%%%%%%%%
%%%%%%%%%%%%%%%%%%%%%%%%%%%%%
%%%%%%%%%%%%%%%%%%%%%%%%%%%%%
%%%%%%%%%%%%%%%%%%%%%%%%%%%%%

The orbital resonance models for QPOs proposed by Abramowicz \& Kluzniak \citep{AbramowiczKluzniak2001, AKST} are particularly based on resonance between epicyclic frequencies which are excited at a well defined \emph{resonance radius} $r_{p:q}$ given by the condition 
\begin{equation}
\frac{\omega_\theta}{\omega_\mathrm{r}}\,\left(a,\,r_{p:q}\right)=\frac{p}{q},
\label{rescondition}
\end{equation}
where $p\,$:$\,q$ is 3:2 in the case of \emph{parametric resonance},  and arbitrary rational ratio of two small integral numbers (1, 2, 3\dots) in the case of \emph{forced resonances}. Notice that in the case of arbitrary forced resonance also the combinational (``beat'') frequencies could be observed including the 3:2 ratio \citep{AbramowiczKluzniak2001, TAKS}.
Such resonance radii are monotonically decreasing functions of  the rotational parametr $a$ (see Fig. \ref{Ratios} - left panel). Resulting resonant frequencies are given generally as a linear combination of epicyclic frequencies at $r_{p:q}$. In \citet{TAKS} it is reported that the resonant frequencies (both observed frequencies, the \emph{upper} and the \emph{lower}) are not monotonic function of $a$ for the case of  3:1 and 5:1 
(5:2) forced resonance models, while for other resonance model ratios (e.g, 2:1, 3:2) discussed in \citet{TAKS}, it is a monotonic function of the rotational parameter $a$.

For the observational consequences, it is important to determine the limiting value of the frequency ratios $\omega_\theta/\omega_\mathrm{r}$$\,=\,$$p/q$, which separates the monotonic and nonmonotonic dependence of the resonant frequencies on the rotational parameter.

Indeed, this monotonicity of some resonant frequencies is resulting from nonmonotonic character of the epicyclic frequencies.
It is known that the radial epicyclic frequency has a local maximum at $r_\mathrm{max(r)}\,$$\equiv\,$$\mathcal{R_\mathrm{r}}$ for $a\in (-1,\,1)$ and its value $\nu_{\mathrm{r}\,(\mathrm{max})}\,(a)$ is increasing with the rotational parameter (see Fig. \ref{KeplerianVsEpicyclic} - right panel). Moreover, outside its maxima it is monotonically decreasing with the radius. From the left panel of Fig. \ref{Ratios} we conclude that $r_{p:q}$(a) must be a \emph{monotonicaly decreasing function} of $a$ [if the horizontal line representing some ratio $p$:$q$ is fixed, then this figure implies necessarily monotonically decreasing function $r_{p:q}(a)$]. Because of this, the resulting resonant frequency (which is just multiple of the radial frequency) must be monotonically increasing for $r_{p:q}$ located outside (or at) the maximum of the radial epicyclic frequency. 
For Schwarzschild black holes the ratio between the epicyclic frequencies at the radius of maximal radial frequency is exactly $\nu_\theta/\nu_\mathrm{r}\,$=$\,2$ ($x\,$=$\,8$) and then it is slighty changing with the rotational parameter growing, reaching the value $\nu_\theta/\nu_\mathrm{r}\,$$\sim\,$1.8 for extremely rotating Kerr black hole ($a\,$=$\,1$).
This gives the limit in the sense that for $p$:$q\,$$>$$\,1.8$, the radius $r_{p:q}$ is surely located above the locations of the maximum of $\nu_\mathrm{r}$.
\\
On the other hand, an analogical consideration shows that $\nu_\theta$ (or $\nu_\mathrm{r}$, if the resonance condition (\ref{rescondition}) is satisfied) is surely decreasing with the rotational parameter, if  $r_{p:q}$ is located under the location of the maximum of $\nu_{\theta}$. The ratio of the epicyclic frequencies at the maximum of $\nu_{\theta}$ is shown on the middle panel of  Fig. {\ref{Ratios}}.  Its minimum is reached for extremely rotating Kerr black holes at $\nu_\theta/\nu_\mathrm{r}\,(r_\mathrm{max})\,$$\sim$$\,2.18$. It means that for black holes the resonance is surely nonmonotonic if $p/q\,$$\geq$$\,2.18$.

It is clear from discussion presented above that the limit for  nonmonotonicity must be located between the values of $p/q$$\,\in\,$$(1.8,\,2.18)$. We checked numerically the loci of eventuall nonmonotonicity and find that the limit is very close to the upper value, i.e., nonmonotonicity of the function  $\nu_\theta/\nu_\mathrm{r}\,(r_\mathrm{p:q},\,a)$ in dependence on $a$ is relevant for forced resonances with
\begin{equation}
p:q\geq 2.18. 
\end{equation}

Figure \ref{Ratios} (right panel) illustrates this limit by examples of  behaviour of $\nu_\theta\,(a, r_{p:q})$ for three different forced resonances which embody the nonmonotonicy (in the sense described above).

\section{Some important implications for the discoseismology of Kerr naked singularities}
\label{Implications}

Without going into details, we expose on simple examples that physics of the wave propagation and oscillations in thin accretion discs around the Kerr naked singularities is considerably different in comparison with the Kerr black hole case.

\subsection{Modified propagation of inertial-acoustic and inertial-gravity waves}

%%%%%%%%%%%%%%%%%%%%%%%%%%%%%%%%
%%%%   Figure  ModesBH                   %%%%%%%%%%%%%
%%%%%%%%%%%%%%%%%%%%%%%%%%%%%%%%
\begin{figure*}[!]
\begin{minipage}{1\hsize}
{\includegraphics[width=\hsize]{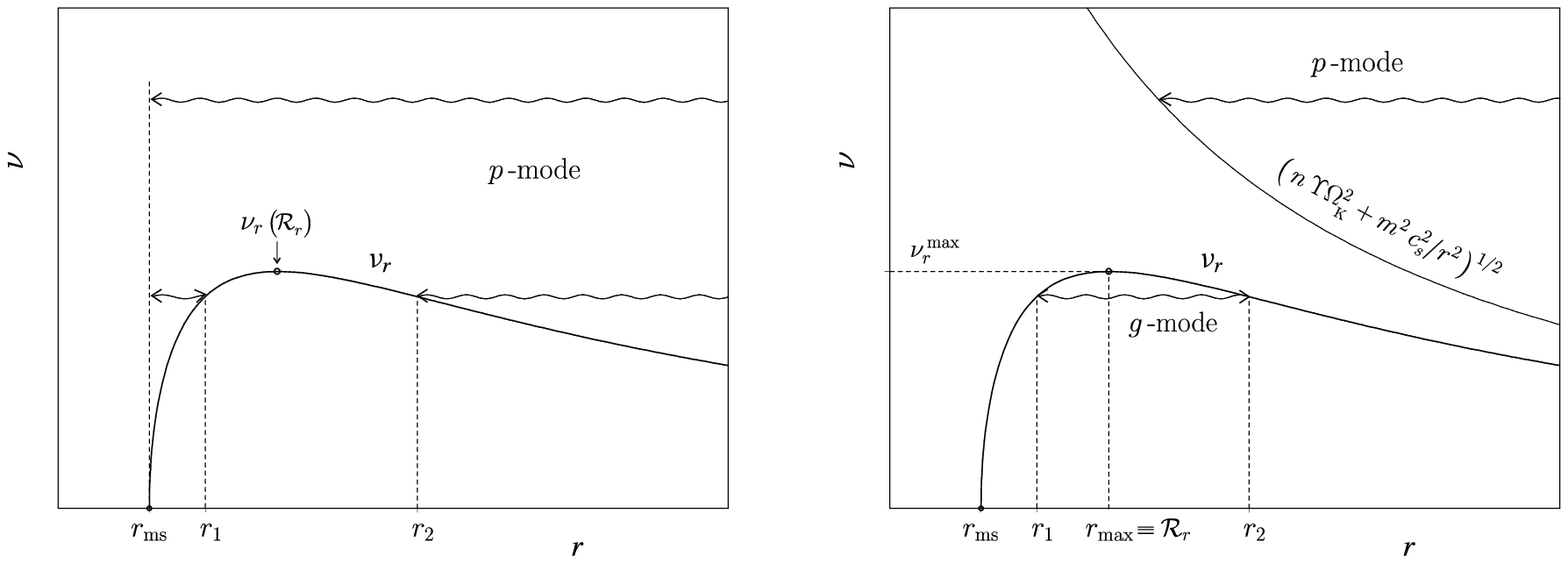}}
\end{minipage}
\caption{\citep[After][]{BlueBook} The simplest examples of wave propagation in black--hole thin accretion disc: for  fundamental axisymmetric mode ($n\,$=$\,m$=$\,0$) see left panel, while the first axisymmetric overtone ($n\,$=$\,1$, $m$=$\,0$) is shown on right one. Both figs are plotted for the Schwarzschild black hole.
%\vspace{.2cm}
}
\label{ModesBH}
\end{figure*}
%%%%%%%%%%%%%%%%%%%%%%%%%%%%%
%%%%%%%%%%%%%%%%%%%%%%%%%%%%%
%%%%%%%%%%%%%%%%%%%%%%%%%%%%%%%%
%%%%   Figure  ModesNaS                   %%%%%%%%%%%%%
%%%%%%%%%%%%%%%%%%%%%%%%%%%%%%%%
\begin{figure*}[!]
\begin{minipage}{1\hsize}
{\includegraphics[width=\hsize]{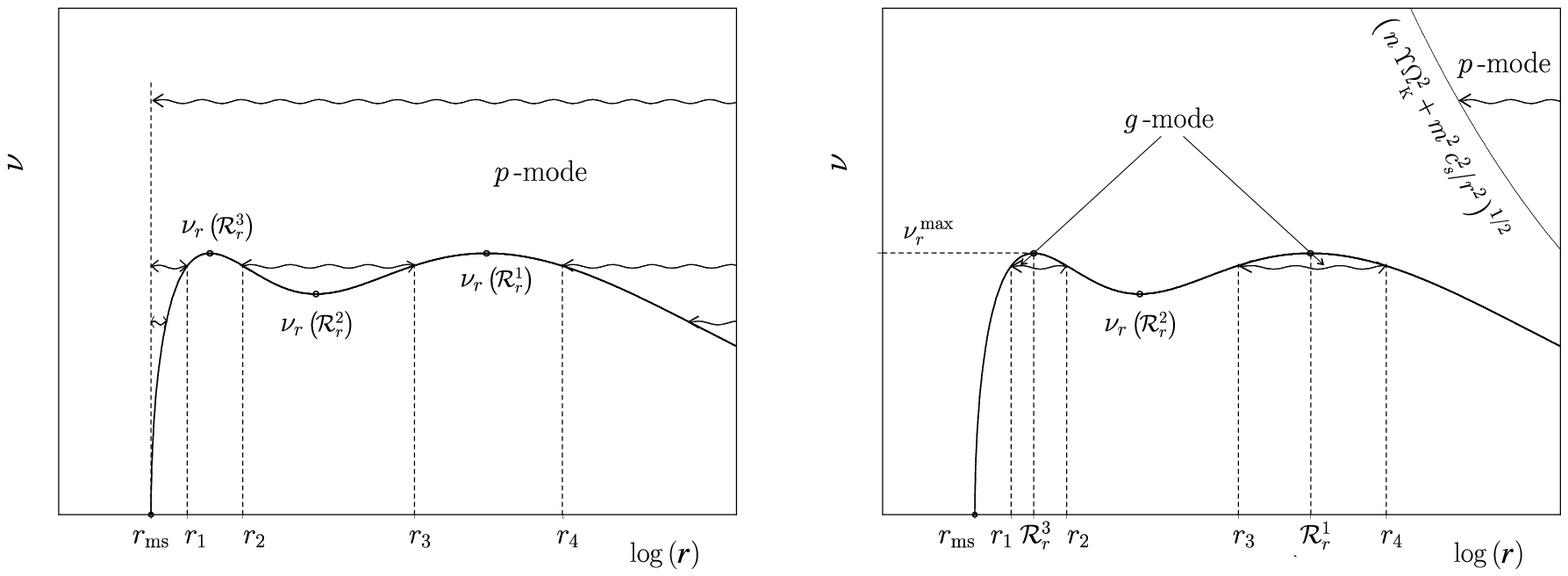}}
\end{minipage}
\caption{The same examples of wave propagation as in Fig. \ref{ModesBH}, but for naked--singularity thin accretion disc (for the simplicity in degenerated case of $a\,$$\doteq$$\,1.012$). Fundamental axisymmetric modes appear on left panel and the first axisymmetric overtone on the right one.
%\vspace{-.1cm}
}
\label{ModesNaS}
\end{figure*}

\noindent
The systematic classification of oscillations and waves in accretion discs according to  the restoring forces acting on them was summarized, e.g., in \citet{BlueBook}.
\smallskip

\noindent
They introduced the following modes for disc oscillations:
\begin{itemize}
\item{The higher frequency modes, i.e., \emph{inertial-acoustic waves}, so called $p$-modes.}
\item{The lower frequency modes, i.e., \emph{inertial-gravity waves}, so called g-modes.}
\item{The inertial-acoustic modes with no node in vertical direction are particularly called the \emph{corrugation waves}.}
\end{itemize}

\noindent
We remind that the wave propagation is given by the dispersion relation \citep{OkazakiKatoFukue1987, Kato1989, NowakWagoner1991, NowakWagoner1992}
\begin{equation}
\left({\omega}^2-\omega_\mathrm{r}^2\right)\left({\omega}^2-n\,\Upsilon\Omega_\mathrm{K}^2-\frac{m^2 c_\mathrm{s}^2}{r^2}\right)={\omega}^2\,k_\mathrm{r}^2c_\mathrm{s}^2,
\label{dispersion}
\end{equation}
where $\omega$ is the angular velocity corresponding to the frequency of  propagated wave, $\Upsilon$ is a correction factor of the order of unity, $k_r$ is the wave vector and $c_\mathrm{s}$ is the vertically averaged sound speed. For the case of axially symmetric ($m\,$=$\,0$) fundamental mode ($n\,$=$\,0$) is this relation reduced to 
\begin{equation}
\omega^2=\omega_\mathrm{r}^2+k_\mathrm{r}^2c_\mathrm{s}^2.
\end{equation}
In this simple case the only waves of $p$-mode ($\omega^2\,$$>$$\,\omega_\mathrm{r}^2$) can  propagate in the radial direction and could be in the black hole case separated into two groups \citep{BlueBook}:
\begin{itemize}
\item{The waves with
\footnote{Here and hencefort in relation to the wave propagation for black holes $\omega_{\mathrm{max}}$ denotes the maximum of $\omega_\mathrm{r}$ at $\mathcal{R}_\mathrm{r}(a)$ (Fig. \ref{RadialExtrems}) in the black-hole range of the rotational parameter.} 
$\omega^2\,$$>$$\,\omega_{\mathrm{max}}^2$
can propagate through the entire region of the disc.}
\item{For the waves with $\omega^2\,$$<$$\,\omega_\mathrm{max}^2$ the propagation region is separated into two discrete portions (see Fig. \ref{ModesBH} - left panel)}:
\begin{itemize}
\item{Inner region $r<r_1$.}
\samepage
\item{Outer region $r>r_2$,}
\end{itemize}
\end{itemize}
\noindent
where $r_1$, $r_2$ are the solutions of  $\omega^2\,$=$\omega_\mathrm{r}^2$; $\pi$-waves with $\omega^2\,$$<$$\,\omega_{\mathrm{max}}^2$ cannot propagate between $r_1$ and $r_2$, however they can propagate outside this region. Because of physical properties of the inner edge of the disc, the $\pi$ waves could be \emph{trapped} in the inner region, i.e., they are reflected between the inner edge and $r_1$ as shown by \citet{KatoFukue1980}.

The overtones ($n\,$$\neq$$0$) in equation (\ref{dispersion}) allow the inertial acoustic $p$-waves as well as inertial gravity $g$-waves. Generally, $p$-waves in this case must have frequency $\omega^2\,$$>$$\,n\,\Upsilon\Omega_{\mathrm{K}}^2$, while for $g$-waves the inequality $\omega^2\,$$<$$\,\omega_\mathrm{r}^2$ must hold.
\\
The $g$-waves which could be trapped (see Fig. \ref{ModesBH} - right panel) in the region around $r_\mathrm{max(r)}$$\,\equiv$$\,\mathcal{R}_\mathrm{r}$ are insensitive to changes of the disc structure, if the condition of moderately geometrically thin disc is satisfied \citep{OkazakiKatoFukue1987, NowakWagoner1992, Perez1997}.

\smallskip

In the case of Kerr naked singularities, the well known picture of the wave propagation, shortly reminded above, is changed due to the different behaviour of the epicyclic frequencies. 
As shown in Sect.~\ref{Properties}, the radial epicyclic frequency has two global maxima for Kerr naked singularities with
$1\,$$<$$\,a\,$$<$$\,a_{\mathrm{c}r}\,$$\doteq$$1.025$.
Thus the propagation of the fundamental axisymmetric ($n\,$$=$$\,0$, $m\,$$=$$\,0$)  $\pi$-waves is changed (of course, the condition $\omega^2\,$$>$$\omega_\mathrm{r}^2$ must be satisfied again):
\begin{itemize}
\item{The waves with
\footnote{In the relation to the wave propagation in discs around Kerr naked singularities, $\omega_{\,\mathrm{MAX}}$ denotes the global maximum of  $\omega_\mathrm{r}$, while $\omega_{\mathrm{max}}$ denotes a local (lower) one, see Fig. \ref{DegeneracyAndRatio}.}~
$\omega^2\,$$>$$\,\omega_{\,\mathrm{MAX}}^2$
can propagate through the entire region of the disc as well as in the black hole case.}
\item{The waves with $\omega^2\,$$<$$\,\omega_{\,\mathrm{MAX}}^2$ could 
be divided into three groups}
\begin{itemize}
\item {For waves with $\omega\,$$>$$\,\omega_\mathrm{max}$ the propagation region splits into two (inner and outer) regions as in the case of black holes.}
\item{For waves with $\omega_\mathrm{min}\,$$<$$\,\omega\,$$<$$\,\omega_\mathrm{max}$ the propagation region splits into three different portions from which two could contain trapped waves.}
\item{For waves with $\omega_\mathrm{min}\,$$>$$\,\omega\,$ the propagation region splits again into inner and outer region as in the black hole case.}
\end{itemize}
\end{itemize}

This modified wave propagation is illustrated in Fig. \ref{ModesNaS} (left panel) for the simple degenerated case $a\,$$\doteq$$1.012$ when $\omega_\mathrm{max}\,$$=$$\,\omega_{\,\mathrm{MAX}}$.

In Fig. \ref{ModesNaS} (right panel) we illustrate that analogically to the situation  described above for trapped $\pi$-waves of fundamental mode, the propagation of trapped $g$-modes ($n\,$$\neq$$\,0$) splits into discrete portions in the case of naked singularities with $a$$\,<\,$$a_{\mathrm{c}r}\,$$\doteq$$\,1.025$.

For the Kerr naked singularities with rotational parameter $a$$\,>\,$$a_{\mathrm{c}r}$, the radial epicyclic frequency has the same properties as in the black hole case and the wave propagation is formally the same as for black holes.

\subsection{Strong resonant frequency}
\label{StrongFrequency}

%%%%%%%%%%%%%%%%%%%%%%%%%%%%%
%%%%%%%%%%%%%%%%%%%%%%%%%%%%%%%%
%%%%   Figure  DegeneracyAndRatio  %%%%%%%%%%%%%
%%%%%%%%%%%%%%%%%%%%%%%%%%%%%%%%
\begin{figure*}[!]
\begin{minipage}{1\hsize}
{\includegraphics[width=\hsize]{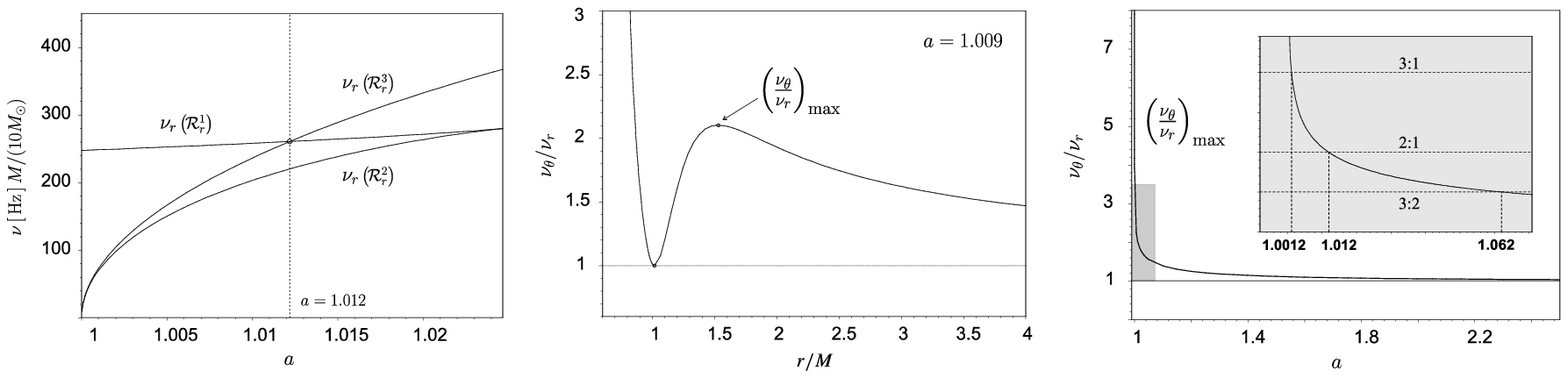}}
\end{minipage}
\caption{Left panel shows the radial epicyclic frequency at particular extrema as function of rotational parameter $a$ between $a\,$= 1 and $a\,$= 1.025. 
We can see that for $a\,$$<$$\,1.012$ the global maximum is situated at $\mathcal{R}_\mathrm{r}^{1}\,(a)$ while for $a\,$$>$$\,1.012$ it is at $\mathcal{R}_\mathrm{r}^{3}\,(a)$. Middle panel exposes the behaviour of  function $\nu_\theta/\nu_\mathrm{r}(x)$ typical for Kerr naked singularities. With increasing parameter $a$
the local maximum of this function is shifted to higher radii. Right panel illustrates that the value of  maximum $\nu_\theta/\nu_\mathrm{r}$ is rapidly decreasing with the rotational parameter growing. Such behaviour of function $\nu_\theta/\nu_\mathrm{r}(x,\,a)$ implies the important consequence for resonance models: in the case of slowly rotating Kerr naked singularities an eventual resonance orbit $r_{p:q}$ (with $p$, $q$ being small integral numbers) is defined unambiguously - the upper limit for this unambiguity  in the case of $p\,$:$\,q$\,$=$ 3:1, 2:1 and 3:2 is denoted.
}
\label{DegeneracyAndRatio}
\end{figure*}
%%%%%%%%%%%%%%%%%%%%%%%%%%%%%
%%%%%%%%%%%%%%%%%%%%%%%%%%%%%
%%%%%%%%%%%%%%%%%%%%%%%%%%%%%

\begin{figure*}[!]
\begin{minipage}{1\hsize}
{\includegraphics[width=\hsize]{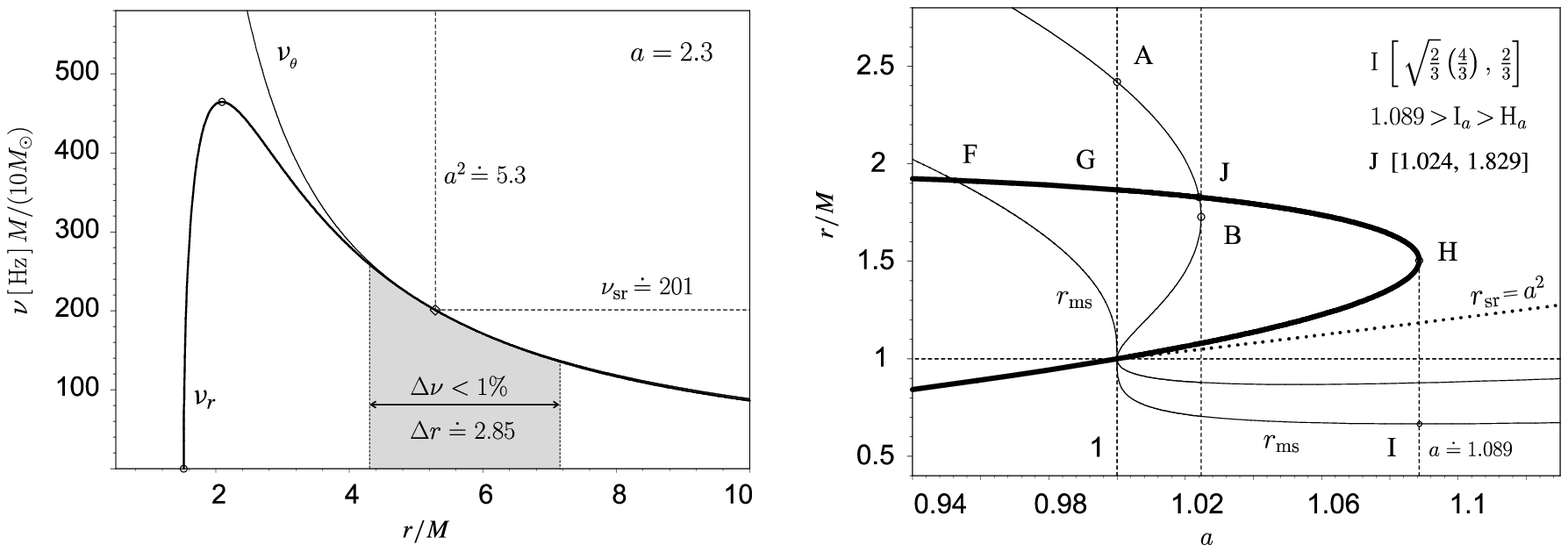}}
\end{minipage}
\caption{Left panel shows the behaviour of epicyclic frequencies for $a\,$=$\,2.3$, the region where frequencies are identical with accuracy of 1\% is denoted. Right panel illustrates the location of the strong resonant frequency (dotted curve $a^2$) in relation to extrema of epicyclic frequencies (thin curve $\mathcal{R}_\mathrm{r}^{k}$, thick one $\mathcal{R}_\theta^{l}$). The critical radius is always located (in radial order) between the first maxima of $\omega_\mathrm{r}$ and the minima of  $\omega_\theta$ (if these exist). The notation of the important points [A\dots J] is in accord with Figs. \ref{RadialExtrems}, \ref{VerticalExtrems}.
%\vspace{.4cm}
}
\label{ExampleAndLocation}
\end{figure*}

It is shown in Sect. \ref{Properties} that for Kerr naked singularities with $a\,$$>$$\,a_{\mathrm{c}\left(\theta \right)}\,$$\doteq$$\,1.089$ the behaviour of epicyclic frequencies is formally similar as for Kerr black holes.
However, for any naked singularity with $a\geq 1$, the epicyclic frequencies (\ref{radial}, \ref{vertical}) can satisfy the equality condition
\begin{equation}
\omega_\mathrm{r}(a,\,x) = \omega_\theta(a,\,x)
\end{equation}
giving \emph{strong resonant phenomenon}
\footnote{Intuitively clear attribution is  well founded in the last subsection \ref{instability}.}
 which occurs at the critical radius
\begin{equation}
\label{resonantradius}
x_\mathrm{sr}=a^2\quad (a\geq 1).
\end{equation}
This means that for any Kerr naked singularity the epicyclic frequency ratio $\nu_\theta/\nu_\mathrm{r}\,(r)$ is a nonmonotonic function which reaches the value 1 at the point given by (\ref{resonantradius}) (see Fig. \ref{DegeneracyAndRatio} - middle panel). The loci of this point is compared with locations of some other important points as show in Fig. \ref{ExampleAndLocation} - the right panel, while the left panel shows an example illustrating radial extension of the strong resonant phenomenon.

%%%%%%%%%%%%%%%%%%%%%%%%%%%%%
%%%%%%%%%%%%%%%%%%%%%%%%%%%%%%%%
%%%%   Figure  ExampleAndLocation  %%%%%%%%%%%%%
%%%%%%%%%%%%%%%%%%%%%%%%%%%%%%%%
%%%%%%%%%%%%%%%%%%%%%%%%%%%%%
%%%%%%%%%%%%%%%%%%%%%%%%%%%%%
%%%%%%%%%%%%%%%%%%%%%%%%%%%%%
%%%%%%%%%%%%%%%%%%%%%%%%%%%%%
%%%%%%%%%%%%%%%%%%%%%%%%%%%%%

Using the relation (\ref{resonantradius}) in (\ref{radial}, \ref{vertical}) we find the {strong resonant frequency} which in terms of the corresponding angular velocity reads
\begin{equation}
\omega_\mathrm{sr}\equiv\omega_\mathrm{r}\,(a,\,a^2) = \omega_\theta\,(a,a^2)=
\left ({{GM_0}\over {r_G^{~3}}}\right )^{1/2}
{\frac {\sqrt {{a}^{2}-1}}{a^
{2} \left( {a}^{2}+1 \right) }},
\end{equation}
and the frequency can be expressed in the form
\begin{equation}
\nu_\mathrm{sr}=\mathrm{32.3\,}\left(\frac{M_\odot}{M}\right)\,{\frac {\sqrt {{a}^{2}- 1}}{{a}^{2} \left(  1+{a}^{2} \right) }}\,\,\mbox{kHz}.
\end{equation}
We note that the strong--resonance phenomenon represents a crucial difference of  Kerr naked singularities from the case of Kerr black hole for which the ratio $\omega_\theta/\omega_\mathrm{r}\,(r)$ is determined as a monotonic function for fixed $a$ (Fig. \ref{Ratios} - left panel). Notice that for Kerr naked singularities with any rotational parameter, the ratio $\omega_\theta/\omega_\mathrm{r}\,$$<$$\,(\omega_\theta/\omega_\mathrm{r})_\mathrm{max}$ being fixed can appear at three different radii
\footnote{This can be important in discussion of observational aspects of resonant phenomena.}. Note that this kind of behaviour is implied by the existence of the strong resonance frequency ($\omega_\theta\,$=$\,\omega_\mathrm{r}$) for any Kerr naked singularity, i.e., it is not restricted to the cases when local extrema of $\omega_\theta$, $\omega_r$ exist.
 
The behaviour of  the epicyclic frequency ratio $\omega_\theta/\omega_\mathrm{r}\,(r)$ typical for Kerr naked singularities is shown in Fig. \ref{DegeneracyAndRatio} (middle panel).
In right panel in Fig. \ref{DegeneracyAndRatio} we plot the value of the local extrema of the ratio $\omega_\theta/\omega_\mathrm{r}$ as a function of the rotational parameter $a$. We see that  for high values of the rotational parameter, the radial and vertical epicyclic frequencies are very close each other in large radial range around $r_\mathrm{sr}$. The example is given in Fig. \ref{ExampleAndLocation} (left panel). 

\begin{figure*}[!]
\begin{minipage}{1\hsize}
{\includegraphics[width=\hsize]{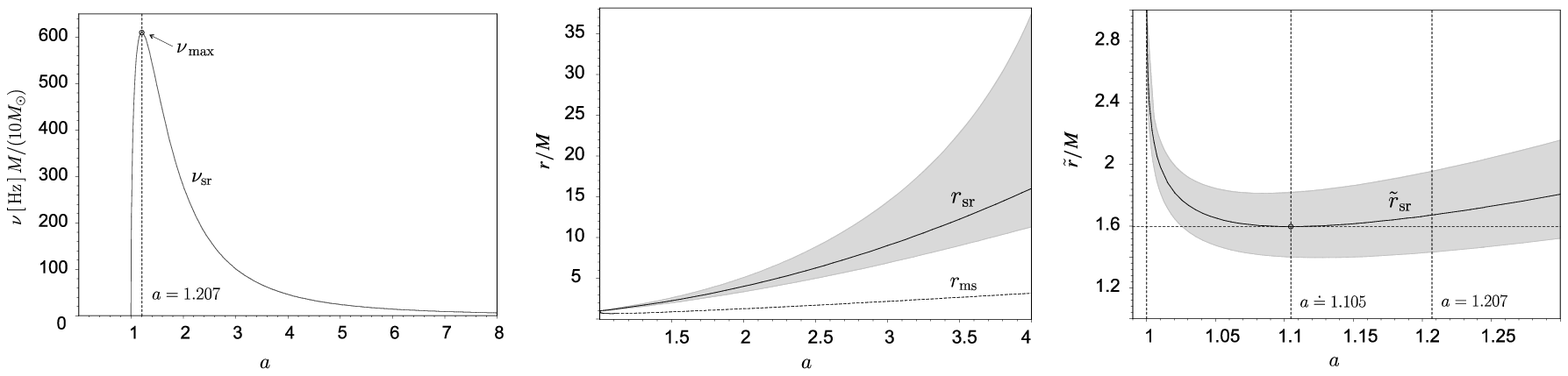}}
\end{minipage}
\caption{The behaviour of strong resonant frequency is shown on left panel. From the middle panel we can see the evolution of  area (around critical radius) where epicyclic frequencies are close, while right panel show the same in the proper distance to marginally stable orbit.}
\label{Confusion}
\end{figure*}

\noindent
We plot the strong-resonance frequency as a function of  the rotational parameter in Fig. \ref{Confusion} (left panel.) The strong-resonance frequency is approaching zero value for extremely rotating Kerr black hole and has a maximum for naked singularities having rotational parameter
\begin{equation}
a_\mathrm{src}\doteq1.207,
\end{equation}
with the corresponding value of  the epicyclic frequency being determined by the relation
\begin{equation}
\label{maximalresonantfrequency}
\nu_\mathrm{src}\doteq 6.1\,\left( \frac{M}{M_\odot}\right)^{-1}\,\mathrm{kHz}.
\end{equation}

In the middle panel of  Fig. \ref{Confusion},  location of the critical radius $x_\mathrm{sr}\,$=$\,a^2$  is shown together with location of the marginally stable orbit and extension of  the \emph {instability}
\footnote{The meaning of the term \emph{instability} used here is discussed in the last subsection \ref{instability}.}
 region of $r$ where the difference between values of the radial and vertical epicyclic frequencies is smaller then 1\%. However, for the values of rotational parameter $a\sim$$1$, it is more convenient to express the region of the disc with 1\% difference of the epicyclic frequencies in terms of the proper radial distance $\tilde{r}$, which has direct physical meaning. There is
\begin{equation}
\tilde{r}=\int_{r_0}^{r_1}\sqrt{g_{\,r\,r}}\,\mathrm{d}r,
\end{equation}
where $g_{\,r\,r}$ denotes the radial metric coefficient of the Kerr metric in the standard Boyer-Lindquist coordinates; the distance is measured from the inner edge of the thin discs, located at $r_\mathrm{ms}$. The result is represented by right panel in Fig. \ref{Confusion}; we have found that location of the strong-resonance frequency is closest to the inner edge of  the Keplerian disc for naked singularity with
\begin{equation}
a_{\,\mathrm{\tilde r}}\doteq 1.105.
\end{equation}
For this value of the rotational parameter the strong resonant frequency is
\begin{equation}
\nu_\mathrm{sr\,in}\doteq 4.3\,\left( \frac{M}{M_\odot}\right)^{-1}\,\mathrm{kHz},
\end{equation}
what is about 70\% of maximum at $a_\mathrm{sr\,max}$ given by (\ref{maximalresonantfrequency}).
However, we remark that the critical radius is always located outside from the innermost part of the disc (see Fig. \ref{ExampleAndLocation}, right panel).

%%%%%%%%%%%%%%%%%%%%%%%%%%%%%
%%%%%%%%%%%%%%%%%%%%%%%%%%%%%%%%
%%%%   Figure  Confusion  %%%%%%%%%%%%%
%%%%%%%%%%%%%%%%%%%%%%%%%%%%%%%%
%%%%%%%%%%%%%%%%%%%%%%%%%%%%%
%%%%%%%%%%%%%%%%%%%%%%%%%%%%%

\subsection{Possible instability of  the accretion disc around Kerr naked singularities}
\label{instability}

The orbital resonance model \citep{KluzniakAbramowicz} demonstrates that {\it fluid accretion flows} admit two linear quasi-incompressible modes of oscillations, vertical and radial, with corresponding eigenfrequencies equal to vertical and radial epicyclic frequencies for free particles. In a particular model of slender torus, the general properties of this modes can be shown: the vertical mode corresponds to a periodic displacement in which the whole torus moves as a rigid body up and down the equatorial plane and each fluid element has a vertical velocity that periodically changes in time, but does not depend on the position. The frequency of the vertical mode is equal to the vertical epicyclic frequency that a ficticious free particle orbiting at the circle of maximum pressure in the torus equilibrium position would have. Behaviour of the radial mode is similar to the vertical one, and in the linear regime these two modes are formally uncoupled. \citet{KluzniakAbramowicz} argue that in the case of more realistic description which includes non-linear effects given by pressure and dissipation, non-linear effects couple the two epicyclic modes what may results in a \emph{resonance}.

One of  such possible resonances, the \emph{parametric resonance}, seems to be most probable explanation of the 3:2 double peak kHz QPOs observed in some galactic microquasars \citep{AKST}. The effect itself is described by the Mathieu equation \citep{LandauLifshitz1976}
\footnote{After denoting the time derivativ $\mathrm{d}/\mathrm{d}t$ by dot.}
\begin{equation}
\delta \ddot \theta + \omega_{\theta}^2\,[ 1 + h \cos (\omega_r t) ]\, 
\delta \theta  =  0,
\end{equation}
that can be formally derived by considering small
deviations of fluid streamlines from planar circular motion governed by a set of  equations \citep{Rebusco2004, Horak2004}
\begin{eqnarray}
\delta \ddot r + \omega_r^2 \, \delta r = 
\omega_r^2 f_r (\delta r, \delta \theta, \delta \dot r, \delta \dot \theta),
\nonumber
\\
\delta \ddot \theta + \omega_{\theta}^2\, \delta \theta = 
\omega_{\theta}^2 f_{\theta} (\delta r, \delta \theta, \delta \dot r, \delta \dot \theta),
\end{eqnarray}
for a particular choice of $f_r$ and $f_{\theta}$, corresponding to
\begin{eqnarray}
\label{Equation4}
\delta \ddot r + \omega_r^2 \, \delta r = 0 ,
\quad
\delta \ddot \theta + \omega_{\theta}^2\, \delta \theta = 
 - \omega_{\theta}^2\, \delta \theta\, \delta r .
\end{eqnarray}
From the theory of the Mathieu equation it is known that  
the parametric  resonance is then excited when
\footnote{We note that the same condition holds for \emph{internal resonance}, which describes systems with {conserved energy} \citep{Horak2004}.}
\begin{equation}
\label{Mequation}
{\omega_r \over \omega_{\theta}} = {\nu_r \over \nu_{\theta}} = 
{2 \over n}, \quad n =1, \,2, \,3 \dots\,.
\end{equation}
The effect is strongest for the smallest possible value of $n$ \citep{LandauLifshitz1976}. Because in the field of  black holes $\nu_r$$\,<\,$$\nu_{\theta}$ (see Sect. \ref{Properties}), the smallest possible value for resonance is $~n = 3$,
i.e, $2\, \nu_{\theta} = 3\, \nu_r$, what explains very well the 3:2 ratio observed in microquasars \citep{AbramowiczKluzniak2004, TAKS}.

As we show in previous text, the point with radial and vertical epicyclic frequency being  equal exists for any Kerr naked singularity. Obviously, at such point the equality (\ref{Mequation}) is satisfied (${\nu_\mathrm{r}}/{\nu_\theta}$$\,=\,$${1}/{1}$$\,=\,$${2}/{n}$; $n\,$=$\,2$),
and the parametric resonance (between the radial and vertical epicyclic frequency) eventually excited at this point is the strongest possible parametric resonance excited between the epicyclic frequencies in the field of Kerr naked singularities. Such 2:2 resonance must be also stronger then the 3:2 parametric resonance in the black hole case \citep{LandauLifshitz1976}.

From this, and from the fact that the radial region with the epicyclic frequencies being nearly equal is rather large, one can expect that at this region both radial and vertical oscillations could be strongly amplified, leading to an instability of the accretion disc
\footnote{Such a claim is motivated by experience from known situations related to the parametric or forced resonance in complex non--linear systems observed in Earth physics \citep{LandauLifshitz1976}. Examples of mathematically possible resonances causing damages of bridges, wings, etc. with no specific physical coupling mechanism known are discussed in \citet{NayfehMook1979}.}.

\bigskip

\section{Conclusions}
\label{conclusions}

For counterrotating Keplerian orbits, properties of the epicyclic frequencies are the same for all Kerr  black holes and naked singularities. Radial epicyclic frequency has always a local maximum, while the vertical epicyclic frequency has no local extrema at $x$$\,>\,$$x_\mathrm{ph}$.

On the other hand, for corotating Keplerian orbits properties of the epicyclic frequencies strongly depend on the rotational parameter of the Kerr spacetimes. Altough, from this point of view, the most important difference between spacetimes with $a$$\,<\,$1 and  $a$$\,>\,$1 is the change of inequality 
%$\omega_\theta>\omega_\mathrm{r}$ to $\omega_\theta\geq\omega_\mathrm{r}$, 
\begin{equation}
\omega_\theta(x)>\omega_\mathrm{r}(x) ~~~ (a<1)
\quad\rightarrow\quad
\omega_\theta(x)\geq\omega_\mathrm{r}(x) ~~~ (a>1),
\end{equation}
we have to distinguish different possibilities according to the existence and relative locations of the local extrema of the epicyclic frequencies.

In the case of  Kerr black holes, the classification according to the properties of the epicyclic frequencies is given in the following way:

\noindent
\begin{itemize}
\item {{BH1}} (Fig. \ref{SummaryBH}a) $\quad$ $       0 < a  < 0.748$\\
$\omega_\mathrm{r} (x,\,a)$ has one local maximum, $\omega_\theta (x,\,a)$ has no local extrema above the photon circular orbit $x_\mathrm{ph}$.

\noindent
\item {{BH2}}
 (Fig. \ref{SummaryBH}b) $\quad$ $0.748 < a  < 0.952$\\
$\omega_\mathrm{r} (x,\,a)$ with one local maximum, $\omega_\theta (x,\,a)$ with one local maximum at $x<x_\mathrm{ms}$.

\noindent
\item{BH3} (Fig. \ref{SummaryBH}c) $\quad$ $0.952 < a  < 1       $\\
$\omega_\mathrm{r} (x,\,a)$ with one local maximum, $\omega_\theta (x,\,a)$ with one local maximum at $x>x_\mathrm{ms}$.
\end{itemize}
In all the cases, the function $(\omega_\theta/\omega_\mathrm{r})\,(x,\,a)$ which is relevant for resonant effects, has a monotonic (descending) character (Fig. \ref{SummaryBH}). Therefore, for a given rotational parameter $a$, there is only one radius allowed for any $p\,$:$\,q$ resonance. However, we have shown that the resulting resonant frequencies are nonmonotonic functions of $a$ for  $p:q>2.18$, what could embarrass eventual spin estimate given by some resonance models. In addition, the curve $\mathcal{R}_\theta(a)$  has a local maximum at $a\doteq0.852$ and we can conlude that the functions $\omega_\theta (x,\,a_1)$, $\omega_\theta (x,\,a_2)$ with $a_1$, $a_2$ being fixed and higher then $a\,$ $\simeq$$\,0.85$  cross each other - this can be also of observational interest.

In the case of Kerr naked singularities, the classification is given in the following way:

\noindent
\begin{itemize}
\item {NaS1}
 (Fig. \ref{SummaryNaS}a) $\quad$ $1 < a  < 1.012       $\\
$\omega_\mathrm{r} (x,\,a)$ has two local maxima and one local minimum between the maxima;\\ 
$\omega_\mathrm{r\,(max)}\,(x_\mathrm{r\,(in)},\,a) < \omega_\mathrm{r\,(max)}\,(x_\mathrm{r\,(out)},\,a)<\omega_{\theta\,\mathrm{(max)}}$,
where $x_\mathrm{r\,(in)}\equiv \mathcal{R}_\mathrm{r}^3,\,x_\mathrm{r\,(out)}\equiv \mathcal{R}_\mathrm{r}^1$.\\
$\omega_\theta (x,\,a)$ has one local minimum and one local maximum.
There is $x_{\theta\,\mathrm{(max)}}\equiv \mathcal{R}_{\theta}^1<x_{\mathrm{r\,(out)}}$.

\noindent
\item {NaS2} (Fig. \ref{SummaryNaS}b) $\quad$ $1.012 < a  < 1.024       $\\
$\omega_\mathrm{r} (x,\,a)$ has two local maxima and a local minimum in between,
\\
$\quad\omega_\mathrm{r\,(max)}\,(x_\mathrm{r\,(out)},\,a) < \omega_\mathrm{r\,(max)}\,(x_\mathrm{r\,(in)},\,a)<\omega_{\theta\,\mathrm{(max)}}$.\\
$\omega_\theta (x,\,a)$ has one local minimum and one local maximum, with  $x_{\theta\,\mathrm{(max)}}<x_{\mathrm{r\,(out)}}$.

\noindent
\item{NaS3} (Fig. \ref{SummaryNaS}c) $\quad$ $1.024 < a  < 1.025       $\\
The same as in the class NaS3, but with
$x_{\theta\,\mathrm{(max)}}>x_{\mathrm{r\,(out)}}$.

\noindent
\item {NaS4}
%(Fig. \ref{SummaryNaS}d)
$\quad$ $1.025 < a  < 1.047       $\\
$\omega_\mathrm{r} (x,\,a)$ with one local maximum.
$\omega_\theta (x,\,a)$ with one local minimum and one local maximum;   $\omega_{\theta\,\mathrm{(max)}}\geq \omega_{\mathrm{r\,(max)}}$.

\noindent
\item {NaS5} (Fig. \ref{SummaryNaS}d) $\quad$ $1.047 < a  < 1.089       $\\
The same as class NS4, but with 
$\omega_{\theta\,\mathrm{(max)}}< \omega_{\mathrm{r\,(max)}}$.

\noindent
\item {NaS6} (Fig. \ref{SummaryNaS}e) $\quad$ $a  > 1.089       $\\
$\omega_\mathrm{r} (x,\,a)$ with one local maximum,
$\omega_\theta (x,\,a)$ with no local extrema. This class is formally similar to the class BH1, but with the crucial exception of the point $x_\mathrm{sr}$.
\end{itemize}

%\onecolumn

We conclude that the properties of the radial and vertical epicyclic frequencies of the Keplerian motion in the case of Kerr naked singularities differ substantially from the case of Kerr black holes and can have strong observational consequences for both resonant phenomena and stability of  accretion discs around Kerr naked singularities.

While the wave propagation in the case of  oscillations in the discs around Kerr naked singularities is substantially different from the black hole case only for spacetimes with $a\in(1,\,1.025)$, the strong resonance effect can occur for any Kerr naked singularity - the strong resonant frequency of  disc oscillations around Kerr naked singularities arise always at the descending part of  the function $\omega_\mathrm{r}(x,\,a)$ (in vicinity of the local minimum of  $\omega_\theta$, if  this exists), i.e., it is always located above the innermost part of the disc. We stress that this phenomenon represents the strongest parametric resonance between the epicyclic frequencies possible in the field of Kerr naked singularities, stronger than in the case of Kerr black holes. Moreover, the area where the effect occurs is large what could finally have a strong influence on the stability of  the Keplerian disc itself.

It follows from the existence of the strong resonant frequency that the function $(\omega_\theta/\omega_\mathrm{r})\,(x,\,a)$ has the same character for all Kerr naked singularities having one local maximum $(\omega_\theta/\omega_\mathrm{r})_\mathrm{max}$$\,>\,$1
and one local minimum at $(\omega_\theta/\omega_\mathrm{r})_\mathrm{min}\,(x_\mathrm{sr}\,$=$\,a^2,\,a)=1$ which corresponds to the strong resonant frequency.
This implies an important consequence for the other resonant phenomena - namely, in the range of frequencies $(\omega_\theta/\omega_\mathrm{r})_\mathrm{max}(a)\geq\omega_\theta/\omega_\mathrm{r}\geq1$,
the resonant effects with the same rational ratio can occur at three different radii $r_{p:q}$.
Using the behaviour of the function $(\omega_\theta/\omega_\mathrm{r})_\mathrm{max}\,(a)$ (Fig. \ref{DegeneracyAndRatio} - right panel) we can conclude that three radii $r_{p:q}$ could occur in the field of Kerr naked singularities with $a\leq 1.0012$ for $p\,$:$\,q$ $\leq 3:1$,  $a\leq 1.012$ for $p\,$:$\,q$ $\leq 2:1$, and $a\leq 1.062$ for $p\,$:$\,q$ $\leq 3:2$. Clearly, in such situation the same resonant frequency ratios could be induced by very different physical phenomena at different parts of the accretion disc.

\bigskip

\begin{acknowledgements}
We thank Prof. Marek Abramowicz and Ji{\v{r}}{\'{\i}} Hor{\'{a}}k for discussions. 
This work was supported by the Czech grant MSM 4781305903. We also thank the perfect hospitality of Nordita (Copenhagen).

\end{acknowledgements}

\clearpage
\onecolumn
\begin{appendix}
\section{Classification of the Kerr spacetimes (due to epicyclic frequencies)}
\label{SetOfFigures}
\vfill

\noindent
\begin{figure*}[h!]
{\includegraphics[width=1\hsize]{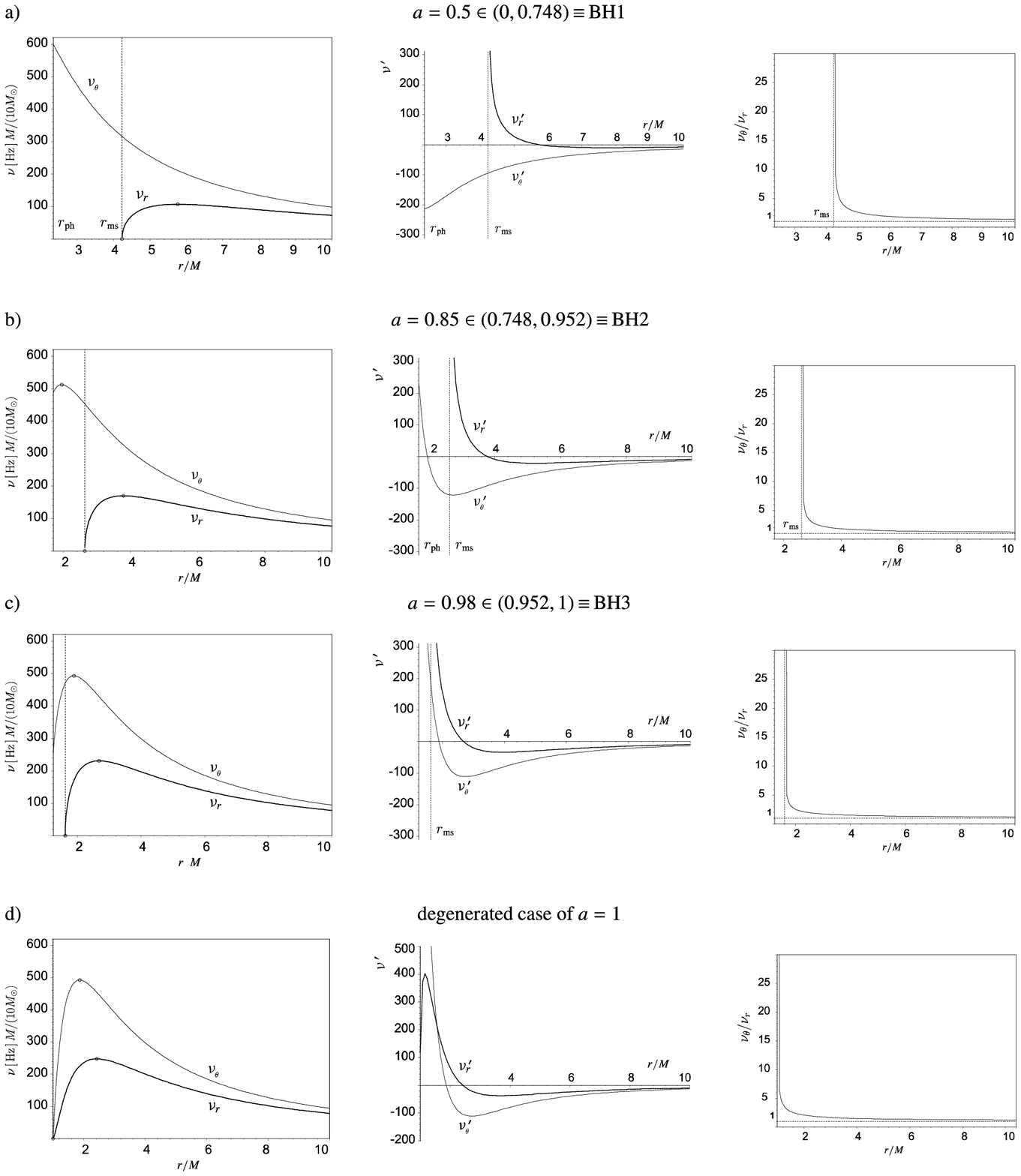}}
\caption{Classification of the Kerr black--hole spacetimes: behaviour of the epicyclic frequencies (left panel), their first derivatives (middle panel), and their ratio $\nu_\theta/\nu_\mathrm{r}$ (right panel) are shown for four representative values of rotational parameter $a$ including the extreme Kerr black hole. Left margin of plots is always situated at the photon circular orbit $r_\mathrm{ph}$, while the marginaly stable orbit $r_\mathrm{ms}$ is denoted by dashed vertical line.}
\label{SummaryBH}
\end{figure*}

\newpage
\begin{figure*}[!]
\vfill
{\includegraphics[width=1\hsize]{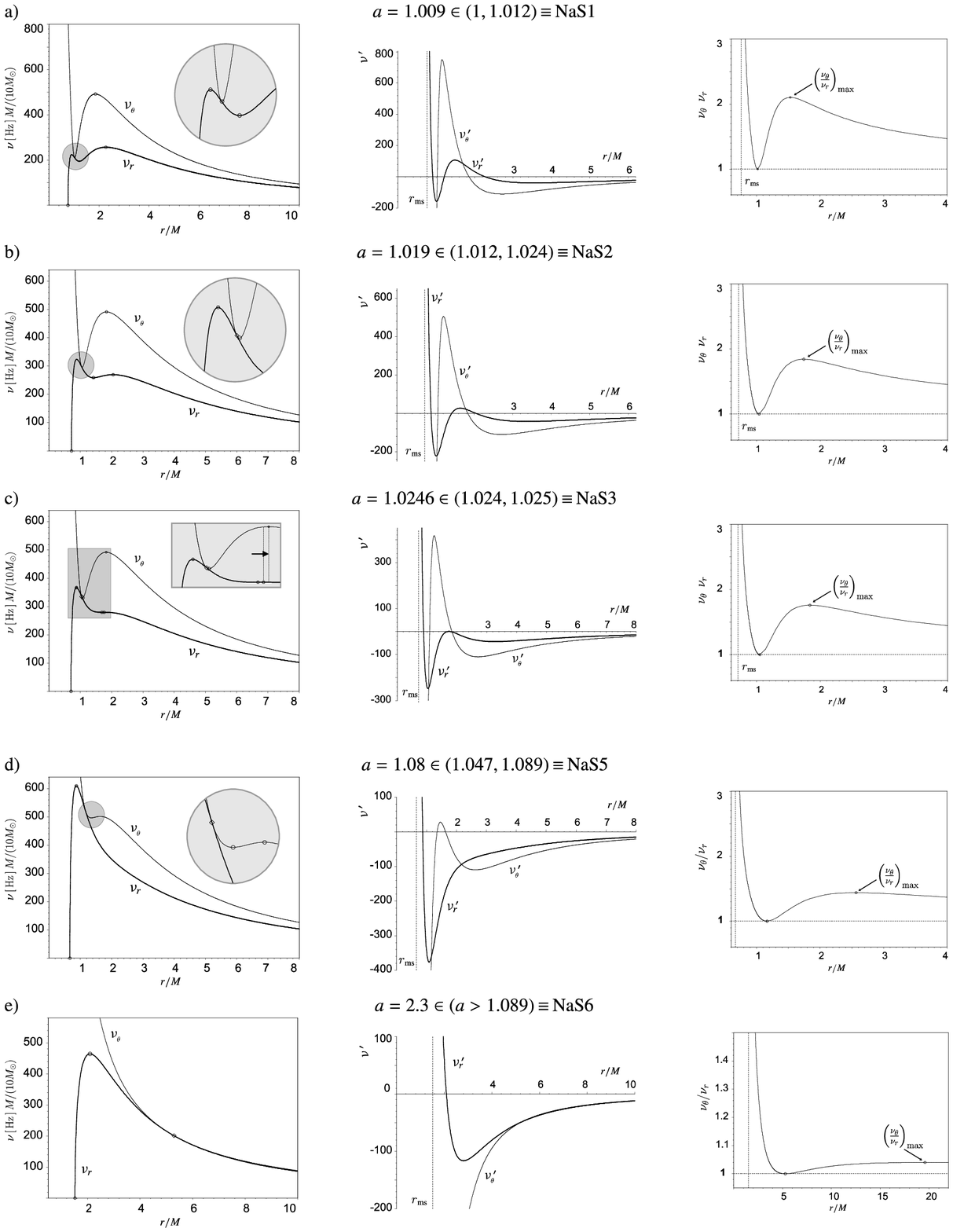}}
\caption{Classification of the Kerr naked--singularity spacetimes: behaviour of the epicyclic frequencies (left panel), their first derivatives (middle panel), and their ratio $\nu_\theta/\nu_\mathrm{r}$ (right panel) are shown for five representative values of rotational parameter $a$, example of the class NaS4 which differs from the class NaS3 by the absence of  the ``radial pair'' maximum--minimum is not shown.}
\label{SummaryNaS}
\end{figure*}

\end{appendix}

\end{document}